\documentclass[a4paper,11pt]{article}
\usepackage{amsmath}
\usepackage{fancyhdr,graphicx}
\usepackage[english]{babel}
\usepackage{hyperref}
\usepackage{rotating} %sidewaystable
\usepackage{multirow}
\usepackage{multicol}        % used for the two-column index
\usepackage{booktabs}
\usepackage{subfig}
\usepackage{relsize}
\usepackage{longtable}
\usepackage{authblk}
\usepackage{color}
\usepackage{array}
\usepackage{amsmath}
    \usepackage[group-separator={,},group-minimum-digits=4]{siunitx} % group separator 
\usepackage{makecell}  
 \usepackage[%
   natbibapa
 ]{apacite}  
\usepackage{tikz}
\usetikzlibrary{matrix,fit}
\usetikzlibrary{calc}
\usetikzlibrary{automata,positioning}
\tikzset{Rectangle2/.style={draw=black!60, fill=black!5, very thick, rounded corners}}
\tikzset{Rectangle2fixed/.style={draw=black!60, fill=black!5, very thick, rounded corners, text width=5em}}
\tikzset{Rectangle2blue/.style={draw=black!60, fill=blue!15, very thick, rounded corners}}
\tikzset{Rectangle2fixedblue/.style={draw=black!60, fill=blue!15, very thick, rounded corners, text width=5em}}
\newcolumntype{P}[1]{>{\centering\arraybackslash}p{#1}}

\usepackage[a4paper,% other options: a3paper, a5paper, etc
        left=1.4cm,
        right=1.4cm,
        top=3.4cm,
        bottom=3.9cm,
]
{geometry}

\title{Epidemics in modern economies} % Complexity Econ and Pandemics

\author[1,2,3]{Torsten Heinrich}
\affil[1]{Faculty for Economics and Business Administration, Chemnitz University of Technology, 09111 Chemnitz, Germany, Email: \href{mailto:torsten.heinrich@wiwi.tu-chemnitz.de}{torsten.heinrich@wiwi.tu-chemnitz.de}}
\affil[2]{Institute for New Economic Thinking at the Oxford Martin School, University of Oxford, Oxford OX1 3UQ, UK}
\affil[3]{Oxford Martin Programme on Technological and Economic Change (OMPTEC), Oxford Martin School, University of Oxford, Oxford OX1 3BD, UK}

\begin{document}

\maketitle
\tableofcontents

\begin{abstract}
% TO.DO: Rewrite abstract
How are economies in a modern age impacted by epidemics? In what ways is economic life disrupted? How can pandemics be modeled? What can be done to mitigate and manage the danger? Does the threat of pandemics increase or decrease in the modern world? The Covid-19 pandemic has demonstrated the importance of these questions and the potential of complex systems science to provide answers. This article offers a broad overview of the history of pandemics, of established facts, and of models of infection diffusion, mitigation strategies, and economic impact. The example of the Covid-19 pandemic is used to illustrate the theoretical aspects, but the article also includes considerations concerning other historic epidemics and the danger of more infectious and less controllable outbreaks in the future.
\end{abstract}

\section{Introduction}
\label{sect:intro}

% TO.DO: introduction
On December 30 2019, Li Wenliang, an ophtalmologist at Wuhan Central Hospital, published some concise facts about a new and dangerous virus in a semi-private chat group: There were multiple cases of SARS (severe acute respiratory syndrome, i.e. severe flu-like symptoms) caused by a coronavirus and they were linked to a particular seafood market; the situation was serious and people should take precautions \citep{Green20}. Li Wenliang was summoned by the police and ordered to not ''spread rumors''. While it is understandable that local authorities did not want to create panic,\footnote{With outbreaks of unknown origin it is difficult to determine the prident way of action. Much of what Chinese authorities did was prudent; they did immediately investigate the seafood market, they quarantined the known patients, they sequenced the virus within three weeks \citep{Zhouetal20}.} it may have helped contain the outbreak. It would be another three weeks before a local lockdown was imposed \citep{Yuanetal20} to curb the spread of the disease. But Li's initial assessment was surprisingly accurate: There was a coronavirus outbreak; the disease now known as COVID-19 has infected more than 150 million people (as of May 2021), claiming more than 3 million deaths around the world (as of May 2021), including Li Wenliang.\footnote{Note that while there are scholars who doubt this explanation of the origins of COVID-19, these are considered fringe opinions and unreliable \citep{Plattoetal20,Michael-Kordatou20,Deslandesetal20,Cohen20}. For a summary of the current state of research, see the official WHO report \citep{Embareketal21}.}

\begin{figure}[tb!]
 \centering
 \includegraphics[width=0.85\textwidth]{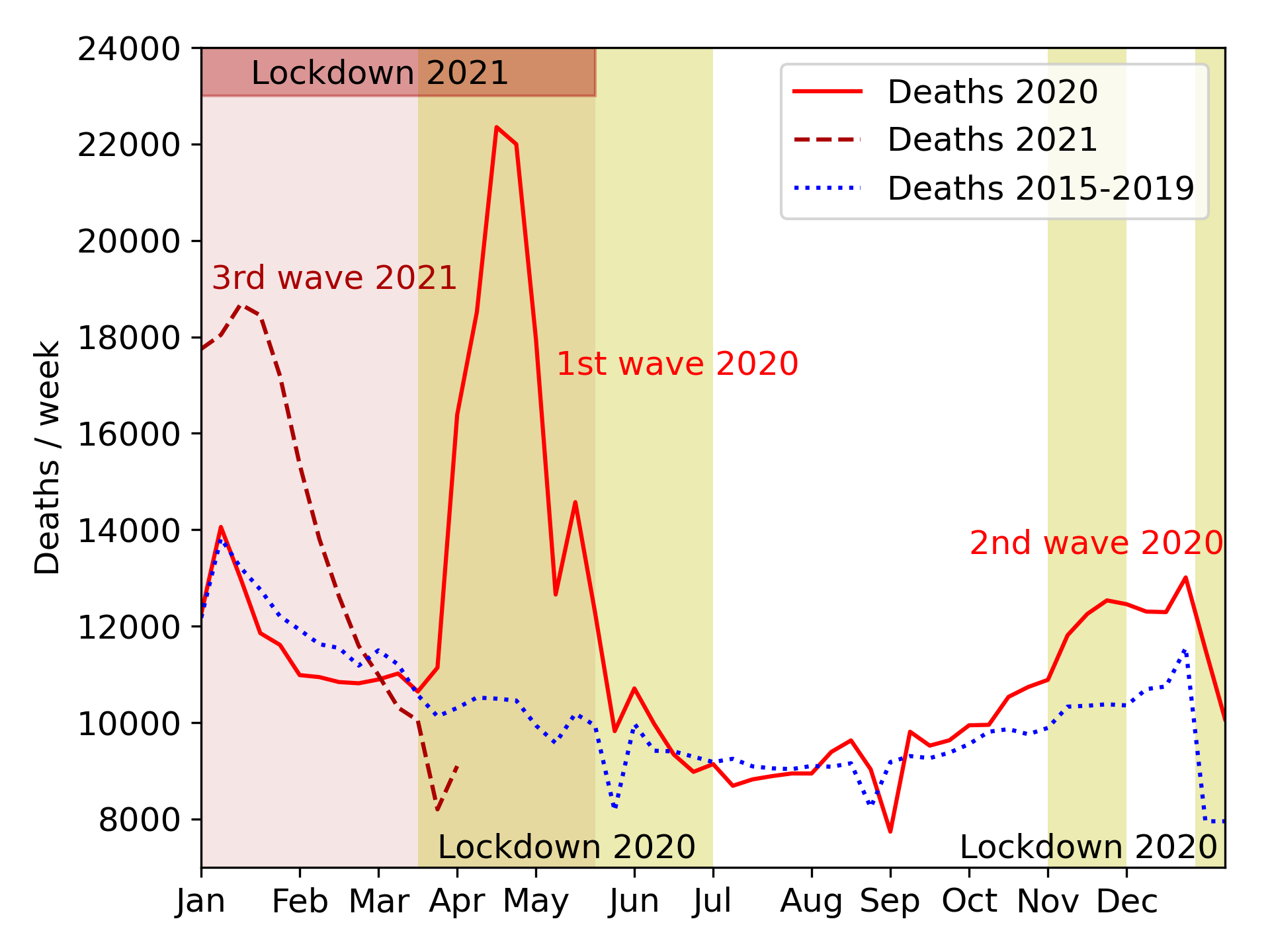}
 \caption{Covid-19 pandemic in England and Wales: Weekly deaths in England and Wales in 2020 and 2021 in comparison to the pre-pandemic average 2015-2019. 2020 lockdown periods are highlighted. (Data from the UK ONS)}
 \label{fig:covid19deathsUK}
 \end{figure}

However, if the lockdown in response to the outbreak in Wuhan was delayed, the reaction in other countries was arguably much worse: Several national leaders denied the existence or severity of the disease in spite of clear evidence to the contrary and in some cases for many months. Initial measures were often half-hearted and epidemiologically nonsensical. In the UK, hundreds of scientists voiced their concerns in an open letter on March 14 2020 \citep{Arrowsmithetal20}: Measures were insufficient and the intensive care capacity of the health care system was going to be exceeded. Stronger measures were taken when it became obvious that this assessment was correct. By now, the scale of the consequences of the pandemic, likely exacerbated by the slow and indecisive response, are becoming clear. The death toll is substantial (Figure~\ref{fig:covid19deathsUK}), economic growth is taking a severe hit, and it is unclear when the pandemic will be defeated.

Why, then, were we so ill-prepared? Why were the measures and the international coordination not more decisive? Why is there still debate about what measures should be taken and about what the current outlook is?

Understanding pandemics requires comprehensive models - social interactions, transport networks and interaction structure have to be considered along with economic systems and the genetics and ecology of the pathogen. Existing empirical data is limited and social and economic systems are varied and change quickly. Tweaking a few parameters in an off-the-shelf model will simply not do. 

One field, that has the capacity and the scope to tackle this problem is, however, complex systems science. Complexity scientists have long advocated \citep{Cincottietal12,Farmeretal12} the use of interdisciplinary approaches, leveraging micro-data, and methods that aim to reflect the complexity of the real world as far as is computationally feasible. 

% TO.DO: Overview over this article
Section \ref{sect:hist} provides a historical perspective, Section~\ref{sect:stopping} offers insights on how pandemics can be managed and stopped. Economic impacts are considered in Section~\ref{sect:econ}. An overview of modeling techniques - compartmental models such as SIR, agent-based models (ABM), and mean field approaches - is given in Section~\ref{sect:modeling}. Section~\ref{sect:conclusion} concludes.

\section{A short history of epidemics}
\label{sect:hist}

Pathogens are older than humanity itself and so are, in all likelihood, pandemics. Throughout recorded history, they have repeatedly swept entire continents or, in the last few centuries, the entire world. Starting with reports of outbreaks some 3350 years ago \citep{Trevisanato07} and early medieval records of what may or may not have been the bubonic plague, examples reach to the black death and modern pandemics of smallpox, cholera and influenza right to the present Covid-19 pandemic. Data for earlier times is sketchy, the record is incomplete until modern times, but the body of evidence is nevertheless substantial. It yields some clear and general facts:

(1) We know what devastating pandemics can do to human society. The black death killed between a third and two thirds of the European population \citep{Alfani/Murphy17}. The diseases introduced in the Americas by conquistadores may have killed up to 92\% of the population - which disrupted the cohesion of societies and the pre-contact civilizations were easily conquered or simply collapsed \citep{Livi-Bacci06,Kochetal19}. 
In Europe, the black death in the 14th century may have forced deep social changes, e.g. through better wages because of labor shortages, and may this have inadvertendly contributed to the rise of Europe. However, it is alco clear that the immediate effects were negative by all accounts - collapse of the irrigation system that had been the core of agriculture in Egypt, decline of income in Spain such that pre-black-death levels were probably only exceeded in the 1800s, in Eastern Europe it may have contributed to the establishment of serfdom \citep{Alfani/Murphy17}. 

(2) Advances sanitation and in medical technology, including vaccines, antibiotics and antiviral drugs, and other treatments have been highly effective in containing epidemics \citep{Madhavetal17,Morensetal08,Morabia20}. Smallpox has been eradicated, cholera, bubonic plague and many other diseases that caused devastating pandemics at some point in history are now treatable and could probably be eradicated given sufficient funding for such an effort. New pandemics have continued to arise, however. Not all of them are under control and new ones continue to emerge. Many scholars have warned of the dangers new pandemics pose to growth, stability, peace, and human society as such \citep{Morensetal08,Cincottietal12,Goldin/Mariathasan14,PastoreyPionttietal19}.

(3) The behavior of pandemics depends crucially on the transmission network, that is population density, social practices, and mobility \citep{Morensetal08,Saunders-Hastings/Krewski16,Lietal18,Clemens/Ginn20,Piret/Boivin21}.\footnote{The effect may not be homogeneous though. \citet{Morens/Fauci07} claim that the effect of mobility has not changed substantially since the 1918-1921 flu pandemic.} Interaction networks have highly asymmetric, sometimes heavy-tailed, degree distributions. That is, the number of contacts is not even; some people have substantially more than most others and are thus both more likely to catch and to spread a given disease. With modern means of transportation, social networks in modern societies also have and the small-world property\footnote{I.e. distant parts of the network are sufficiently connected for dispersion of the pandemic to be possible and rapid.}, meaning the epidemics will not be contained in local populations. Thus the properties of the interaction network will dominate that of outbreaks the size distribution of which will be equally asymmetric \citep{Newman02}. There are indications that the outbreak size may in some cases be heavy-tailed; \citet{Fukui/Furukawa20} find this for Covid-19 for instance.

(4) While increased mobility may have contributed to the spread and endemicity (continued presence in a host population) of pandemic diseases \citep{Morensetal08,Piret/Boivin21}, there were only a handful of pandemics in the last century that health systems could not efficiently and quickly contain, most notably the H1N1 virus (Spanish influenza) of 1918-1921 ($\geq$ 50 million deaths \cite{Morensetal08}), HIV/AIDS ($\approx$ 30 million deaths, 75 million infections so far), and COVID-19 ($\approx$ 3 million deaths, 150 million infections so far).

(5) There are other potential determinants of the emergence of pandemics such as sanitation, the state of public health, and environmental disruption \citep{Morensetal08,Piret/Boivin21}.

(6) Unchecked, any epidemic will spread such that the number of cases grows exponentially until it reaches the capacity of the susceptible population \citep{DArienzo/Coniglio20}. The doubling peropd may be quite short; less than a week in unchecked Covid-19 for instance \citep{Muniz-Rodriguezetal20,LaMaestraetal20}. As a result, the timing of public health measures is crucial. A week may make the difference between an insignificant local outbreak and the near-collapse of the health care system with tens of thousands of deaths. In the case of Covid-19, the slow pace of the reaction in many countries (cf. \citep{Arrowsmithetal20}) made the crisis much worse. There is often a latency period during which subjects may be contagious without realizing they are infected, thus exacerbating the spread of the disease. Symptomless but contagious cases - a hallmark of Covid-19 outbreaks - have the same effect. 

\section{Stopping epidemics}
\label{sect:stopping}

It has been known for a very long time that epidemics somehow spread between people and that modifying the transmission network with measures like quarantines is effective in curbing the spread of the diseases. El Amarna letter EA 96, written about 3350 years ago describes this: ''I will not permit the entrance of the men of Simyra [int]o my city; there is a pestilence in Simyra.''\footnote{The letter is written by a local chief to the governor of Byblos, Rib-Haddi, whose position is described in the quote above. The author of the letter expresses outrage about this quarantine policy and demands it to be lifted.} \citep{Youngblood62}. Quarantine became standard procedure in pre-modern times and long before it was understood how pandemics are caused \citep{Morabia20}. However, it was often difficult to enforce and extensive lockdowns remained prohibitively expensive even for the richest countries until a few decades ago. 

The COVID-19 pandemic constitutes the first opportunity for us to deploy modern technology and stop a global pandemic in its tracks, although we have not succeeded so far. Thanks to epidemiology, network science, and also complex systems science, we have detailed models of the development of the present pandemic - comprehensive studies with constructive policy recommendations on every aspect of the crisis emerged fairly quickly after the start of the pandemic \citep{Fergusonetal20,Pichleretal20,delRioChanonaetal20}. It is possible to send large sections of the labor force to work from home - especially in developed economies far more than half of the labor force works jobs in the tertiary (service) sector, some of which can be done remotely. Global travel networks can be scaled down while maintaining communications via modern communications technologies and enabling logistics and global supply chains to continue, many tasks have been automated anyway. Other important aspects of society such as education and leisure activities are also shifting online. What is more, today's economies are not at the brink of collapse in spite of many months of lockdowns. 

Strategies to contain epidemics - not only Covid-19 - usually include the following \citep{Fergusonetal06,Colizzaetal07}:
\begin{itemize}
  \item Reduce the network density via lockdowns or closure of businesses, the hospitality sector etc.
  \item Compartmentalize the network to prevent it from spreading to other countries or communities.
  \item Remove transmission channels, e.g. animal vectors, unsafe water supply, droplets in the air in indoor venues.
  \item Reduce susceptibility by vaccination.
  \item Increase the capacity of the health care sector.
  \item Contain the spread of the epidemic to as few individuals as possible in order to limit the expected number of mutations.
\end{itemize}
All of these measures have been deployed at one point of the other against the Covid-19 pandemic. They all follow the same epidemiological strategy: increase the selection pressure on the virus beyond it's capacity to adapt. A different strategy is possible: increase the mutation rate enough for the replication process to falter causing the loss of functional adaptations and making the virus population less potent (error catastrophe \citep{Eigenetal88}). It has been suggested that this might play a role in drugs used to manage active Covid-19 infections \citep{Jena20}, although SARS-Cov-2 (the virus causing Covid-19) is a positive-stranded RNA-virus. These viruses have a comparatively low mutation rate, especially compared to retroviruses like HIV.  

\section{Epidemics and the economy}
\label{sect:econ}

\begin{figure}[tb!]
 \centering
 \includegraphics[width=0.85\textwidth]{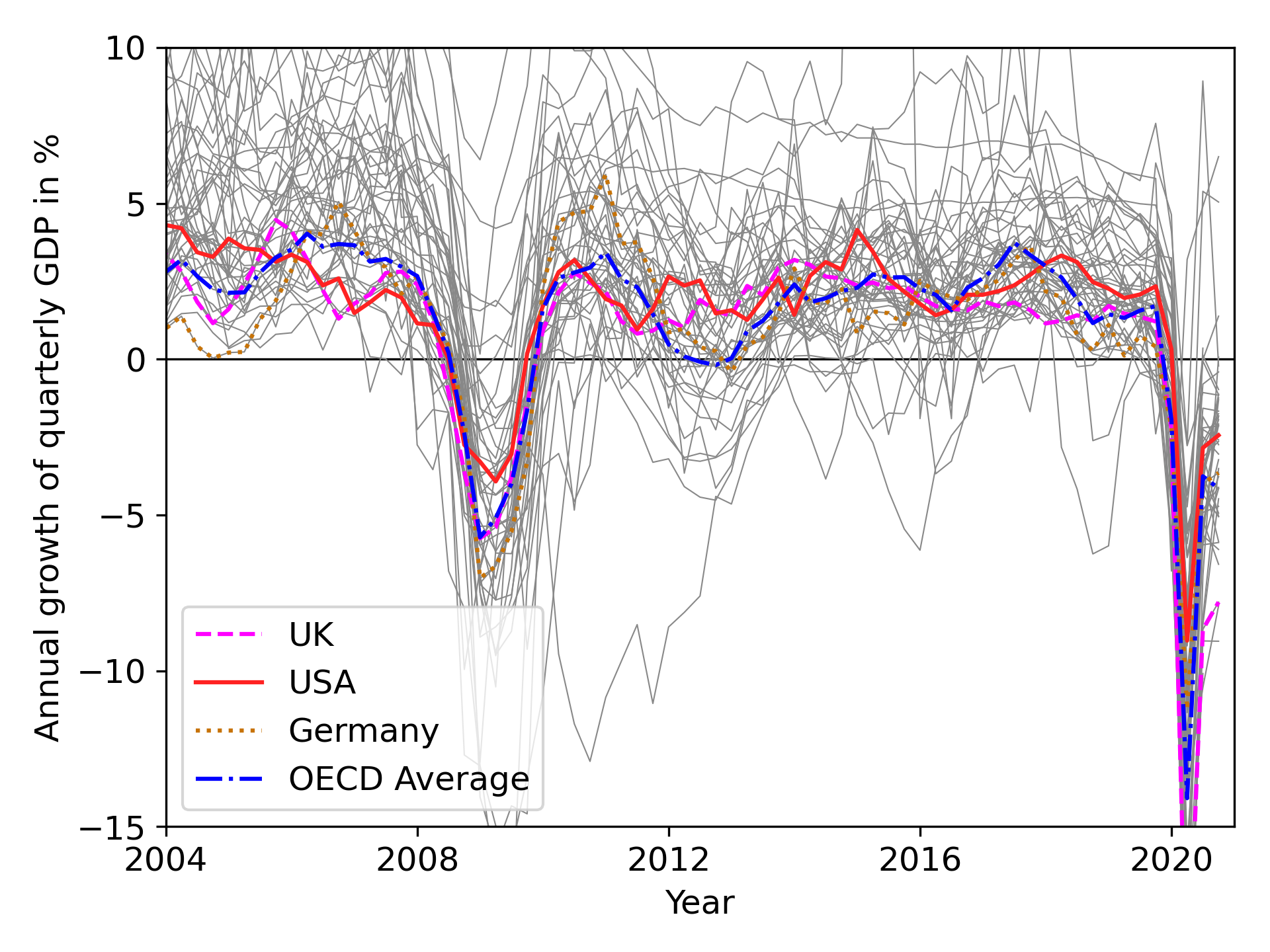}
 \caption{Effect of the Covid-19 pandemic on GDP growth (annual growth rates of quarterly GDP) in OECD countries and other developed economies. (Data from OECD: doi: 10.1787/b86d1fc8-en.)}
 \label{fig:covid19gdp}
 \end{figure}

\begin{figure}[tb!]
 \centering
 \includegraphics[width=0.85\textwidth]{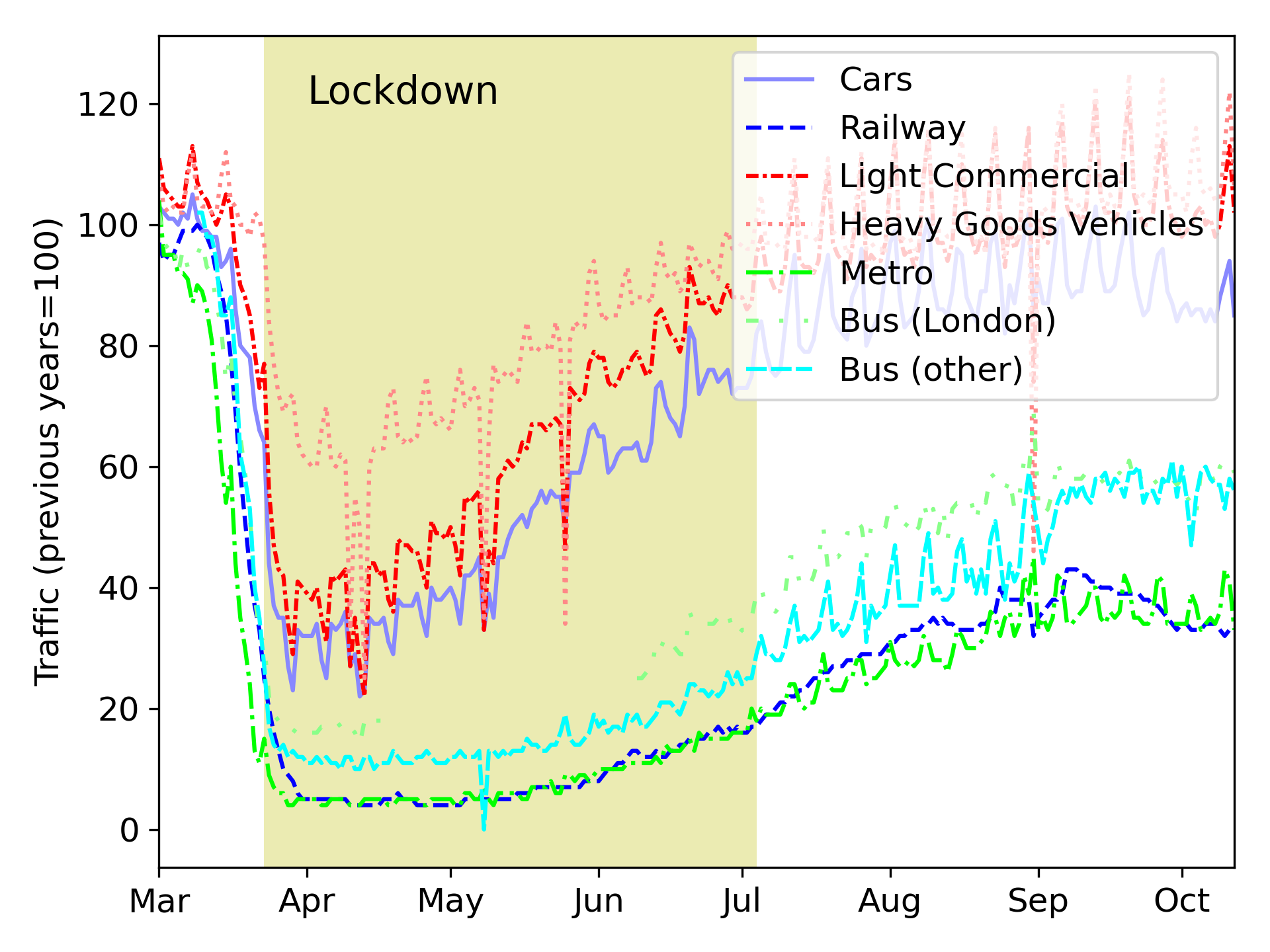}
 \caption{Effect of the lockdown during the 1st wave of the Covid-19 pandemic in the UK: (Data from the UK Department of Transport.)}
 \label{fig:covid19trafficUK}
 \end{figure}

\begin{figure}[tb!]
 \centering
 \includegraphics[width=0.85\textwidth]{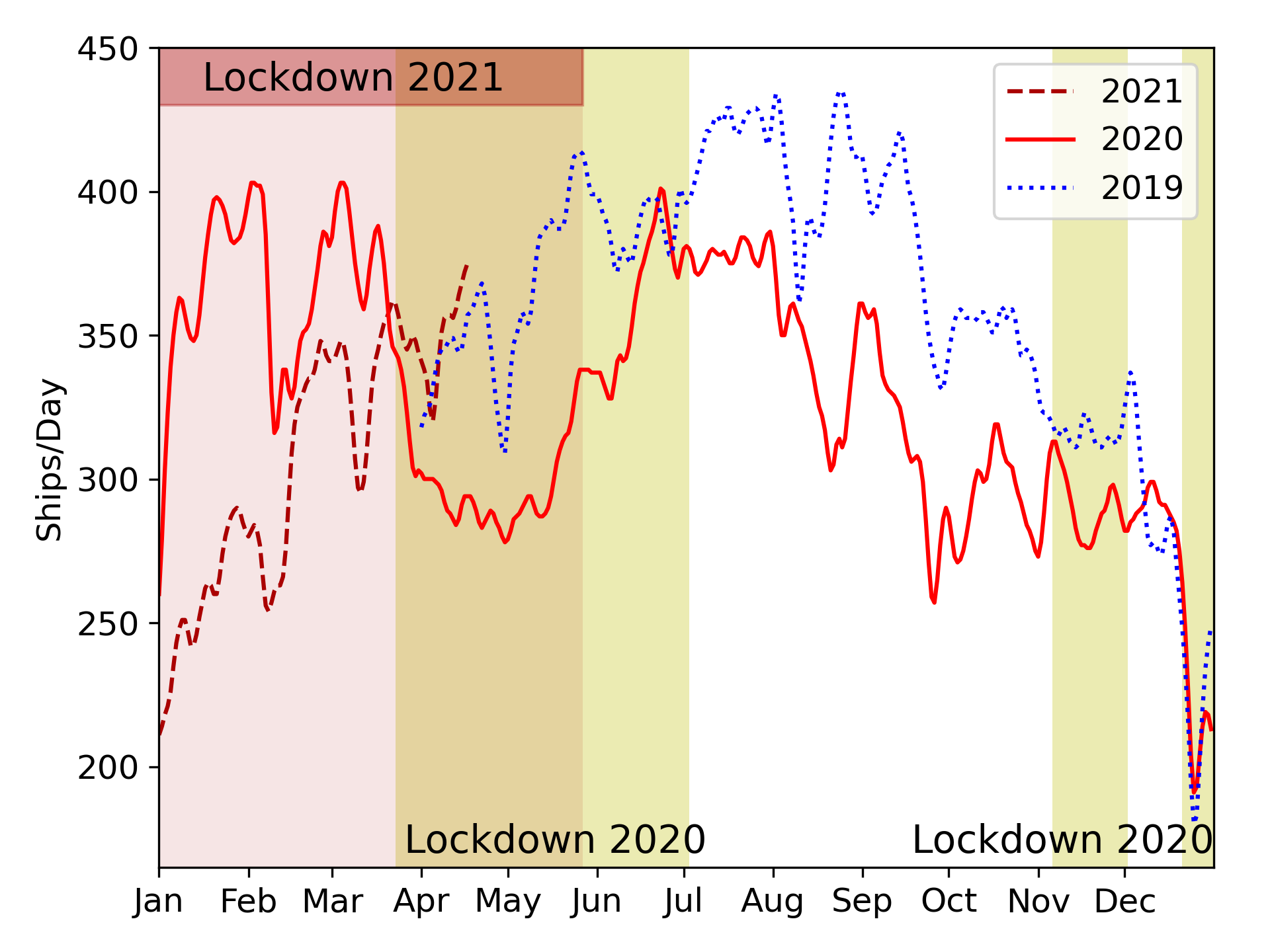}
 \caption{Shipping logistics in the UK during the Covid-19 pandemic: Number of ship visits to UK ports 2019, 2020, 2021. (Data from the UK ONS)}
 \label{fig:covid19shipsUK}
 \end{figure}

\begin{figure}[tb!]
 \centering
 \includegraphics[width=0.85\textwidth]{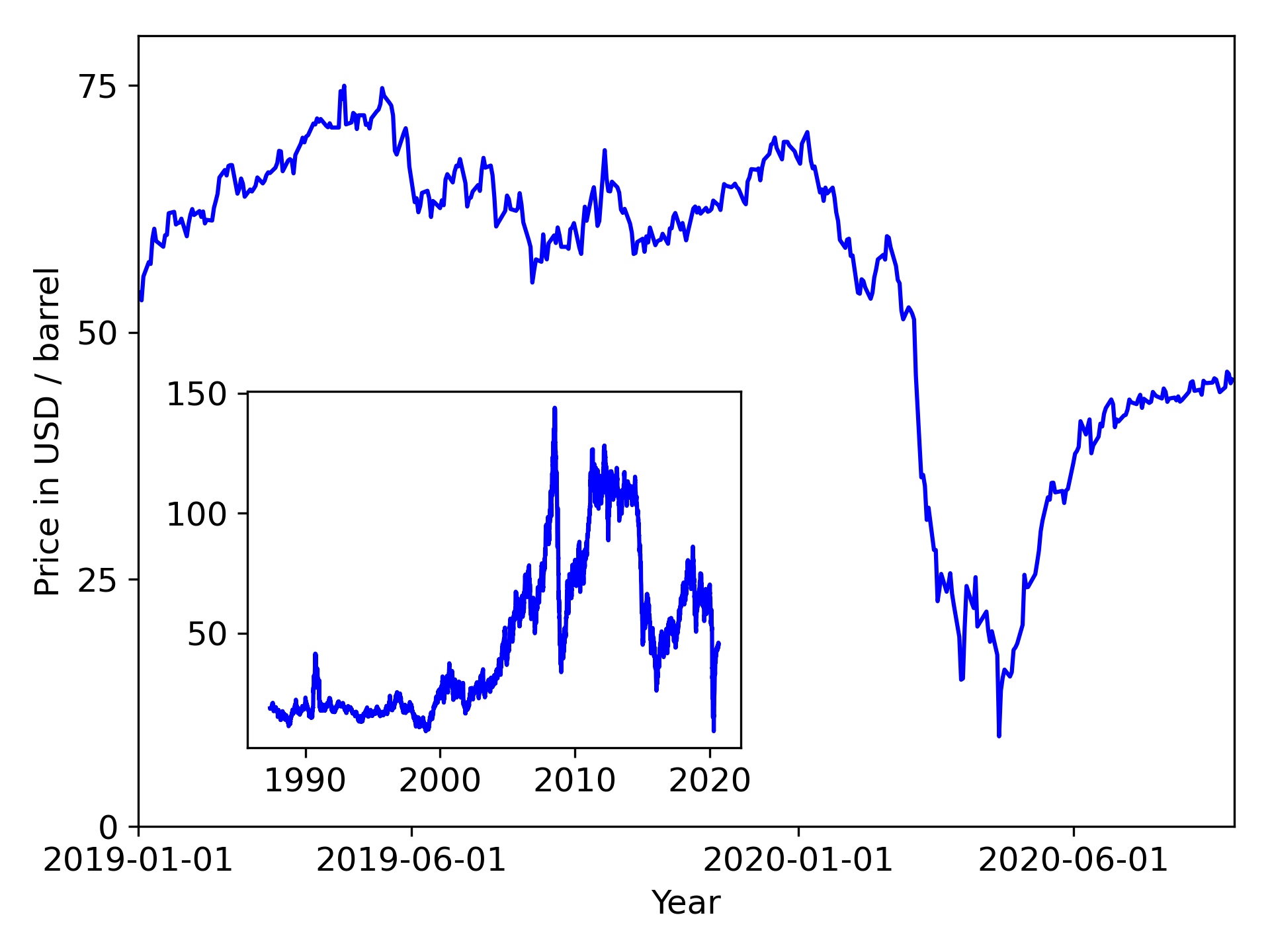}
 \caption{Effect of the Covid-19 pandemic on the price of crude oil. (Data from datahub.io.)}
 \label{fig:covid19oil}
 \end{figure}

The COVID-19 pandemic has changed society and the economy substantially - and even so, in spite of our best efforts, the pandemic is not defeated and may yet lead to further and perhaps devastating consequences. Beyond the immediate effects like the devastating jump in the death rate (Figure~\ref{fig:covid19deathsUK}), the pandemic had effects on all parts of socio-economic life. The GDP collapsed around the globe (Figure~\ref{fig:covid19gdp}), transport logistics ground to a halt (Figures~\ref{fig:covid19trafficUK} and \ref{fig:covid19shipsUK}), there were shortages because of bottlenecks in storage and transportation and later because production of microchips, bicycles and other commodities could not keep up. Meanwhile the price of oil collapsed to practically zero (Figure \ref{fig:covid19oil}), and to negative values for futures. Consumers are saving huge quantities of money. Significant numbers of employees have been laid off or furloughed and parts of the economy, especially in the hospitality sector, are only maintained with government subsidies. The adjustment that will follow when the subsidy schemes will finally end may be as turbulent as the beginning of the pandemic.

% TO.DO: Socio-economic impacts

While the considerations would apply similarly to other pandemics, it is convenient to use the current Covid-19 pandemic as an example. Virtually all other recent pandemics have been brought under control relatively quickly with the exception of the 1918-1921 influenza and the HIV pandemic. The 1918-21 influenza occured in a very different environment (much less sophisticated medical technology, war, social hardship, widespread poverty, different demographic structure) \citep{Morens/Fauci07}; HIV on the other hand progresses and spreads slowly and has thus completely different properties.\footnote{HIV will not as easlily lead to sudden and catastrophic changes in the economy, although the long-term effects in the worst-hit countries are also catastrophing and do not leave the economy unscathed \citep{Canning06}.}

Fairly quickly after the start of the 2020 Covid-19 lockdowns, the question about re-opening safely and mitigating the economic efects emerged. DSGE modelers combined epidemiological SIR models (see below) with standard DSGE models and concluded that the rebound was going to be swift - a V-shaped recovery \citep{Eichenbaumetal20,Acemogluetal20}. Complexity economics on the other hand used data-driven approaches to build models that took - among other things - the sectoral level into account \citep{delRioChanonaetal20,Richiardietal20,Pichleretal20,Sharmaetal20}. Starting from the finding that the impact on different sectors differs widely \citep{delRioChanonaetal20,Richiardietal20}, \citet{Pichleretal20} for instance suggested a partial reopening with consumer-facing sectors remaining closed at first. They use a sectoral agent-based model combined with SIR-model and other aspects and calibrated with the sectoral shocks found by \citep{delRioChanonaetal20}. \footnote{There are many other studies that follow similar, but less sophisticated, approaches, e.g. \citet{Inoue/Todo20,Barrotetal20} or equilibrium models with sectoral levels and input-output matrices but with the pandemic modeled as a simple shock \citep{Baqaee/Farhi20}.} Many dynamic or ABM approaches agreed that a V-shaped recovery was unlikely and variously suggesting W-shaped or L-shaped developments of macroeconomic indicators \citep{Sharmaetal20,Gomez-Pineda20}. Combined with the earliest findings by \citet{Fergusonetal20} that suggested that multiple lockdowns may be necessary to quell repeated outbreaks, and considering the subsequent second and third waves of the pandemic (Figure~\ref{fig:covid19deathsUK}), this finding has been vindicated - there has been no V-shaped recovery so far except possibly for some Asian economies (Figure~\ref{fig:covid19gdp}).

Beyond the development of economic growth, substantial changes can be expected at the micro-level:
\begin{itemize}
  \item With layoffs and uneven impact in different sectors, there are effects on inequality: \citep{Mattanaetal20,Finck/Tillmann20,Casarico/Lattanzio20} agree that moinortities, women, and employees in low-skilled jobs are are disproportionally affected by layoffs. Some authors suggested that stimulus packages may reduce inequality in the short term \citep{Bronkaetal20}, although it is doubtful that this effect will last.
  \item \citet{Haldane20} pointed out that central bank balances are increasingly inflated, more so than at any other point in recent times. It is not clear what the effect of this will be in the mid- to long term.
  \item Substantial skills have been developed in working and collaborating remotely; communication technologies (video conferencing tools in particular) are more sophisticated. However, contrary to popular belief, being forced to work from home still seems less productive than the regular pre-pandemic work-schedule, although this effect was less strong for highly skilled jobs and those who had experience with work from home \citep{Morikawa20}. 
  \item At the same time, a lock-in effect as analyzed in the 1980s by \citet{David85} and \citet{Arthur89} had taken place. Zoom Inc. has emerged as the clear industry leader, followed by Microsoft (Teams), Google (Hangouts) and others. All of these tools have been in the news because of privacy concerns at one point of the other - there are privacy-conscious and distributed alternatives like BigBlueButton and Jitsi, but these are less well-known.
  \item With production interruptions occurring because of outbreaks and lockdowns anyway, it is possible that in some industries the re-investment (upgrade, modernization) cycle will be brought forward. Driven by technological developments that were under way long before the crisis \citep{Brynjolfsson/McAfee15,Brynjolfssonetal17}, this may lead to an earlier replacement of human employees by AI and higher unemployment and vanishing of entire categories of jobs in the wake of the crisis \citep{Pissarides20}.
%  \item 
\end{itemize}

% TO.DO: approaches
%     Fergusonetal20: multiple lds
%     Pichleretal20: sectoral reopening possible - 
%     Eichenbaumetal20,Acemogluetal20: V-shape
%     Sharmaetal20,Pichleretal20,Gomez-Pineda20: W/L shape
%     delRioChanonaetal20,Richiardietal20: strong sectoral effects, substantial numbers of jobs at risk
%     Bronkaetal20: poverty reduction w stimulus package (short term, paper is from june 2020???)
%     Mattanaetal20, Finck/Tillmann20, Cassarico/Lattanzio20: inequality
%     Haldane20
%     Pissarides20
%     Sayed/Peng20
%     Morikawa20
%   
%
% Byambasurenetal20,Fontanet/Cauchemez20
% vs GBD Lourencoetal20

These effects would be expected to occur in a simular way in other pandemics to come, unless the economy, governments, or corporations take steps to prevent undesirable outcomes from this list. 

\section{Modeling epidemics}
\label{sect:modeling}

% TO.DO: SIR etc - \citep{Fergusonetal06,Colizzaetal07}
\subsection{Compartmental models}

One of the simplest and most straightforward models for epidemics, originating over 100 years ago,\footnote{The earliest contribution in this field was \citet{Ross16}; the classical SIR model was suggested by \citet{Kermack/McKendrick27}. They are commonly used today \citep{Fergusonetal06,Colizzaetal07}.} are the compartmental models \citep{DArienzo/Coniglio20}. The population is divided into a finite number of groups characterized by their current infection status: susceptible $S$, infected $I$, recovered/immune $R$, dead $D$. The transitions between these groups are then modeled dynamically with a set of ordinary differential equations:

\begin{equation}
  \begin{array}{r c l}
    \frac{dS}{dt} & = & - \beta IS\\
    \frac{dI}{dt} & = & \beta IS - \gamma I\\
    \frac{dR}{dt} & = & \gamma I\\    
  \end{array}
\label{eq:sir}
\end{equation}

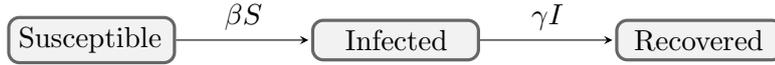
\begin{figure}[tb!]
 \centering
\begin{tikzpicture}%
  [>=stealth,
   shorten >=1pt,
   node distance=1cm,
   on grid,
   auto,
   every node/.style={}
  ]
\node[Rectangle2fixed]       (S)                        {Susceptible};
\node[Rectangle2fixed]       (I)   [right=4.cm of S]    {\hspace*{0.3cm}Infected};
\node[Rectangle2fixed]       (R)   [right=4.cm of I]    {\hspace*{0.1cm}Recovered};
\path[->]
    (S) edge[bend right=0] node {$\beta S$} (I)
    (I) edge[bend right=0] node {$\gamma I$} (R)
   ;
\end{tikzpicture}  
 \caption{Simple SIR model.}
 \label{fig:SIR}
 \end{figure}

$\beta$ is the rate of infection for each pairing of a susceptible and an immune infividual; $\gamma$ is the rate at which infected persons recover. The model makes the simplification that all persons are identical except for their infected status and that interactions between all individuals are equally likely. As a result, the development of the outbreak can easily be computed.  
Figure~\ref{fig:SIR} visualizes the intuition behind this model. 

Compartmental SIR models can easily be extended. They could include deaths (SIRD models) include limited immunity, i.e. reversion from immune to susceptible (SIRS models), and many other aspects. They can also be included in other, more comprehensive models (as in \citep{Pichleretal20,Eichenbaumetal20}) or extended to allow for non-trivial interaction networks (see below).

Like in real-life epidemics, the outbreak will show an close-to-exponential growth until it approaches the capacity boundary - i.e., as long as $S$ remains high. Many public health efforts, including many of those in response to the Covid-19 pandemic, are designed to react to the $R_0$ number. This number is the exponential growth rate

\begin{equation}
  R_0 = \frac{\beta}{\gamma}.
\end{equation}

While it is easy to work with the $R_0$ number in a model environment such as this one, where everything is known, the number is more difficult to measure in real outbreaks.

% TO.DO: ABM
\subsection{Agent-based models}

Instead of operating at the aggregated level the transition rates $\beta I$ and $\gamma$ can be interpreted as transition probabilities $\widetilde{\beta_i}$, $\widetilde{\gamma_i}$ at the level of the individual agent. Equation~\ref{eq:sir} becomes

\begin{equation}
\begin{array}{r c l}
    \frac{dp(S_i)}{dt} & = & - \widetilde{\beta_i} p(S_i)\\
    \frac{dp(I_i)}{dt} & = & \widetilde{\beta_i} p(S_i) - \widetilde{\gamma_i} p(I_i)\\
    \frac{dp(R_i)}{dt} & = & \widetilde{\gamma_i} p(I_i)\\    
  \end{array}.
\label{eq:sir:abm}
\end{equation}

and $p(S_i)$, $p(I_i)$, and $p(R_i)$ are now probabilities that agent $i$ is in the respective state.  

Computationally, this can be simulated by shifting each agent from state to state with probabilities $\widetilde{\beta_i}$ and $\widetilde{\gamma_i}$ in each time period. Even with modest computation power, huge models of this type can be simulated \citep{Bagnietal02}. The advantage is that agents can now be heterogeneous and there is no need to assume a complete network (where agery agent could infect every other agent) - any network structure can be studied. Further, adding additional states such as latent infections, co-morbidities, etc. no not make the simulation substantially more complex. Of course, in this case a simulation approach is used - with all advantages and drawbacks of that approach (for an overview, see, e.g., \citet{Graebneretal19}).

% TO.DO: SIR mean fields
\subsection{Mean fields in compartmental models}

Another way to allow for heterogeneity and non-trivial network structures is to solve the agent-based version of the compartmental model (Equation~\ref{eq:sir:abm}) as a mean-field model \citep{Sharkey08}.

As above, all transmission risks for agent $i$ are folded into $\widetilde{\beta_i}$: 

\begin{equation}
  \widetilde{\beta_i} = \sum_{j=1}^{N} \mathcal{T}_{ji} \frac{p(I_j,S_i)}{p(S_i)}
  \label{eq:sir:abm:beta}
\end{equation}

where $\mathcal{T}$ is the matrix of the transmission rates, $N$ is the number of agents, $p(S_i)$ is the probability that $i$ is susceptible and $p((I_j,S_i))$ is the joint probabilities that $i$ is susceptible and $j$ is infectious.

With statistical independence $p(I_j,S_i)=p(I_j)p(S_i)$, we obtain (after substituting \ref{eq:sir:abm:beta} into \ref{eq:sir:abm})

\begin{equation}
\begin{array}{r c l}
    \frac{dp(S_i)}{dt} & = & - p(S_i) \sum_{j=1}^{N} \mathcal{T}_{ji} p(I_j) \\
    \frac{dp(I_i)}{dt} & = & p(S_i) \sum_{j=1}^{N} \mathcal{T}_{ji} p(I_j) - \widetilde{\gamma_i} p(I_i)\\
    \frac{dp(R_i)}{dt} & = & \widetilde{\gamma_i} p(I_i)\\    
  \end{array}.
%\label{eq:sir:abm}
\end{equation}

The expected numbers of agents in each state are

\begin{equation}
\begin{array}{r c l}
    S & = & \sum_{i=1}^{N} p(S_i) \\
    I & = & \sum_{i=1}^{N} p(I_i) \\
    R & = & \sum_{i=1}^{N} p(R_j) \\    
  \end{array}.
%\label{eq:sir:abm}
\end{equation}

Expectations for $p(S_i)$, $p(I_i)$, and $p(R_i)$ are

\begin{equation}
\begin{array}{r c l}
    p(S_i) & = & \frac{S}{N} \\
    p(I_i) & = & \frac{I}{N} \\
    p(R_i) & = & \frac{R}{N} \\
  \end{array}.
%\label{eq:sir:abm}
\end{equation}

At this point, we can drop $R$ and $\frac{dR_i}{dt}$, because the system is closed and $S+I+R=N$ must hold at every time step.

We obtain development equations 

\begin{equation}
\begin{array}{r c l}
    \frac{dS}{dt} & = & - \sum_{i=1}^{N} \sum_{j=1}^{N} \mathcal{T}_{ji} p(I_j) p(S_i) = - \frac{\sum_{j=1}^{N} \mathcal{T}_{ji}}{N^2} IS\\
    \frac{dI}{dt} & = & \sum_{i=1}^{N} \sum_{j=1}^{N} \mathcal{T}_{ji} p(I_j) p(S_i) - \sum_{i=1}^{N} \widetilde{\gamma_i} p(I_i)= \frac{\sum_{j=1}^{N} \mathcal{T}_{ji}}{N^2} IS - \sum_{i=1}^{N}\widetilde{\gamma_i} \frac{I}{N}\\
  \end{array}.
\label{eq:sir:mf}
\end{equation}

Comparing Equation~\ref{eq:sir:mf} to \ref{eq:sir} shows that the equivalent of $\beta$ here is $\frac{\sum_{j=1}^{N} \mathcal{T}_{ji}}{N^2}$ while that of $\gamma$ is $\frac{\sum_{i=1}^N\widetilde{\gamma_i}}{N}$. Hence, we also have 

\begin{equation}
  R_0 = \frac{\frac{\sum_{j=1}^{N} \mathcal{T}_{ji}}{N^2}}{\frac{\sum_{i=1}^N\widetilde{\gamma_i}}{N}}.
\end{equation}

Note that these expressions can be simplified if we assume homogeneous recovery rates $\widetilde{\gamma_i}=\gamma$ and homogeneous transmission probabilities. If we have homogeneous transmission probabilities $\beta$ on all existing connections in the network $\mathcal{T}=\beta\mathcal{A}$ (where $\mathcal{A}$ is the simple adjacency matrix and $n=||\mathcal{A}||=\frac{\sum_{i+1}^N\sum_{j=1}^N\mathcal{A}}{N}$ is the average degree; the average number of neighbors), we get 

\begin{equation}
\begin{array}{r c l}
    \frac{dS}{dt} & = & - \frac{\beta n}{N} IS\\
    \frac{dI}{dt} & = & \frac{\beta n}{N} IS - \gamma I\\
  \end{array},
\end{equation}

which reduces back to Equation~\ref{eq:sir} for the case of a complete network, where everyone is everyone else's neighbor, because then $n=N$.

\subsection{Modeling the impact on the economy}

Compartmental models or their extensions can be used to model the development of a pandemic, especially its scale. This has been used as an input in economic models either in the form of a shock \citep{Baqaee/Farhi20} or by including the SIR-model as such \citep{Pichleretal20,Acemogluetal20,Eichenbaumetal20}. The latter approach has the advantage that the impact of policy decisions and economic activity on the course of the pandemic can be incorporated in the SIR-based ABM or mean field. 

The economic part of the model will often concentrate on the sectoral level \citep{Pichleretal20,Baqaee/Farhi20} conveniently modeled using input-output-tables. The impact of the pandemic on the labor force - possible labor shortages in particular sectors or possible layoffs is also often included \citep{delRioChanonaetal20,Acemogluetal20}. 

ABMs are particularly versatile in this context. Compared to simpler models which separate the two aspects - the epidemiological and the economic model - integrated ABMs can offer important insights into feedback effects, differences between sectors, and possible systemic risks for both the expected future development and hypothetical scenarios (for alternative policy measures for instance).

\section{Conclusion}
\label{sect:conclusion}

% TO.DO: conclusion
% TO.DO: move to conclusion??
Complex systems science and complexity economics have helped enormously in understanding the present Covid-19 pandemic \citep{Fergusonetal20} and its socio-economic consequences \citep{delRioChanonaetal20,Pichleretal20,Sharmaetal20}. Effects on economic growth and the labor market have been modeled in detail. Effects on inequality, productivity, and financial stability are suspected. Many other effects of the Covid-19 crisis on, e.g., innovation, insurance, financial stability, health care economy, etc. are likely, but few findings have been clearly established. 

With vaccinations there is hope that the pandemic will be brought under control soon, i.e. within months, in most countries. This is in line with initial estimates \citep{Byambasurenetal20,Fontanet/Cauchemez20}. However, it is also clear that Covid-19 will not be the last pandemic and SARS-Cov-2 not the last emerging pathogen. Integrated trade networks, improved mobility and higher population densities all contribute to this. In spite of some cautioning voices in recent years (e.g., \citet{Goldin/Mariathasan14}) and good theoretical understanding of how pandemics spread and how they can be stopped, society, policy makers and economic systems have been largely unprepared for what happened. It is crucial that this should not happen again. There are many pathogens that cause periodic localized outbreaks or have the potential to do so - and some of them (Ebola, Marburg-virus, new strains of influenza, multiresistant bacteria?, more virulent HIV?) could be much more devastating than Covid-19. It is important that modern society and public officials learn to trust scientists and researchers - including complex systems scientists and complexity economists - more than political strongmen, who were very vocal in the early stages of the Covid-19 pandemic, and who have agendas and strong opinions, but no expertise.

\bibliographystyle{apacite}
\bibliography{harticle}

\end{document}